\documentstyle[12pt,epsfig]{article}
\def \pbarp {p{\overline p}}

\begin{document}
\begin{flushright}
FERMILAB-Pub-00/311-E \\
\today\\
\end{flushright}
\begin{center}
\begin{Large}
{\bf Measurement of the Two-Jet Differential Cross Section in 
$\pbarp$ Collisions at $\sqrt{s} = 1800$ GeV }

\end{Large}
\end{center}
\font\eightit=cmti8
\def\r#1{\ignorespaces $^{#1}$}
\hfilneg
\begin{sloppypar}
\noindent
T.~Affolder,\r {23} H.~Akimoto,\r {45}
A.~Akopian,\r {38} M.~G.~Albrow,\r {11} P.~Amaral,\r 8 S.~R.~Amendolia,\r {34} 
D.~Amidei,\r {26} K.~Anikeev,\r {24} J.~Antos,\r 1 
G.~Apollinari,\r {11}
% ...
\end{sloppypar}
\vskip .026in
\begin{center}
(CDF Collaboration)
\end{center}

\vskip .026in
\begin{center}
\r 1  {\eightit Institute of Physics, Academia Sinica, Taipei, Taiwan 11529, 
Republic of China} \\
\r 2  {\eightit Argonne National Laboratory, Argonne, Illinois 60439} \\
\r 3  {\eightit Istituto Nazionale di Fisica Nucleare, University of Bologna,
I-40127 Bologna, Italy} \\
\r 4  {\eightit Brandeis University, Waltham, Massachusetts 02254} \\
\r 5  {\eightit University of California at Davis, Davis, California  95616} \\
\r 6  {\eightit University of California at Los Angeles, Los 
Angeles, California  90024} \\  
\r 7  {\eightit Instituto de Fisica de Cantabria, CSIC-University of Cantabria, 
39005 Santander, Spain} \\
\r 8  {\eightit Enrico Fermi Institute, University of Chicago, Chicago, 
Illinois 60637} \\
\r 9  {\eightit Joint Institute for Nuclear Research, RU-141980 Dubna, Russia}
\\
\r {10} {\eightit Duke University, Durham, North Carolina  27708} \\
\r {11} {\eightit Fermi National Accelerator Laboratory, Batavia, Illinois 
60510} \\
\r {12} {\eightit University of Florida, Gainesville, Florida  32611} \\
\r {13} {\eightit Laboratori Nazionali di Frascati, Istituto Nazionale di Fisica
               Nucleare, I-00044 Frascati, Italy} \\
\r {14} {\eightit University of Geneva, CH-1211 Geneva 4, Switzerland} \\
\r {15} {\eightit Glasgow University, Glasgow G12 8QQ, United Kingdom}\\
\r {16} {\eightit Harvard University, Cambridge, Massachusetts 02138} \\
\r {17} {\eightit Hiroshima University, Higashi-Hiroshima 724, Japan} \\
\r {18} {\eightit University of Illinois, Urbana, Illinois 61801} \\
\r {19} {\eightit The Johns Hopkins University, Baltimore, Maryland 21218} \\
\r {20} {\eightit Institut f\"{u}r Experimentelle Kernphysik, 
Universit\"{a}t Karlsruhe, 76128 Karlsruhe, Germany} \\
\r {21} {\eightit Center for High Energy Physics: Kyungpook National
University, Taegu 702-701; Seoul National University, Seoul 151-742; and
SungKyunKwan University, Suwon 440-746; Korea} \\
\r {22} {\eightit High Energy Accelerator Research Organization (KEK), Tsukuba, 
Ibaraki 305, Japan} \\
\r {23} {\eightit Ernest Orlando Lawrence Berkeley National Laboratory, 
Berkeley, California 94720} \\
\r {24} {\eightit Massachusetts Institute of Technology, Cambridge,
Massachusetts  02139} \\   
\r {25} {\eightit Institute of Particle Physics: McGill University, Montreal 
H3A 2T8; and University of Toronto, Toronto M5S 1A7; Canada} \\
\r {26} {\eightit University of Michigan, Ann Arbor, Michigan 48109} \\
\r {27} {\eightit Michigan State University, East Lansing, Michigan  48824} \\
\r {28} {\eightit University of New Mexico, Albuquerque, New Mexico 87131} \\
\r {29} {\eightit The Ohio State University, Columbus, Ohio  43210} \\
\r {30} {\eightit Osaka City University, Osaka 588, Japan} \\
\r {31} {\eightit University of Oxford, Oxford OX1 3RH, United Kingdom} \\
\r {32} {\eightit Universita di Padova, Istituto Nazionale di Fisica 
          Nucleare, Sezione di Padova, I-35131 Padova, Italy} \\
\r {33} {\eightit University of Pennsylvania, Philadelphia, 
        Pennsylvania 19104} \\   
\r {34} {\eightit Istituto Nazionale di Fisica Nucleare, University and Scuola
               Normale Superiore of Pisa, I-56100 Pisa, Italy} \\
\r {35} {\eightit University of Pittsburgh, Pittsburgh, Pennsylvania 15260} \\
\r {36} {\eightit Purdue University, West Lafayette, Indiana 47907} \\
\r {37} {\eightit University of Rochester, Rochester, New York 14627} \\
\r {38} {\eightit Rockefeller University, New York, New York 10021} \\
\r {39} {\eightit Rutgers University, Piscataway, New Jersey 08855} \\
\r {40} {\eightit Texas A\&M University, College Station, Texas 77843} \\
\r {41} {\eightit Texas Tech University, Lubbock, Texas 79409} \\
\r {42} {\eightit Istituto Nazionale di Fisica Nucleare, University of Trieste/
Udine, Italy} \\
\r {43} {\eightit University of Tsukuba, Tsukuba, Ibaraki 305, Japan} \\
\r {44} {\eightit Tufts University, Medford, Massachusetts 02155} \\
\r {45} {\eightit Waseda University, Tokyo 169, Japan} \\
\r {46} {\eightit University of Wisconsin, Madison, Wisconsin 53706} \\
\r {47} {\eightit Yale University, New Haven, Connecticut 06520} \\
\r {(\ast)} {\eightit Now at Carnegie Mellon University, Pittsburgh,
Pennsylvania  15213}
\end{center}

\renewcommand{\baselinestretch}{2}
\large
\normalsize

\begin{center}
{\bf Abstract}
\end{center}
A measurement is presented of the two-jet differential 
cross section, $d^3\sigma/dE_T d\eta_1 d\eta_2$, at 
center of mass energy $\sqrt{s} = 1800$ GeV in $\pbarp$ collisions.
The results are based on an integrated luminosity of 86~${\rm pb}^{-1}$
collected during 1994-1995 by the CDF collaboration at the 
Fermilab Tevatron collider. 
The differential cross section is measured as a function of the 
transverse energy, $E_T$, of a jet in the pseudorapidity region
$0.1 < |\eta_1| < 0.7$ for four different pseudorapidity bins  
of a second jet restricted to $0.1 < |\eta_2| < 3.0$.
The results are compared with next-to-leading order QCD calculations 
determined using the CTEQ4 and MRST sets of parton distribution functions. 
None of the sets examined in this analysis provides a good description of the data. 

\vspace*{0.5in}

\clearpage

Jet production in proton-antiproton collisions results predominantly 
from hard interactions between two initial state partons. Theoretical 
developments in both perturbative next-to-leading order (NLO) and 
parton shower Monte Carlo calculations 
permit calculation of many QCD jet processes with theoretical 
uncertainties small enough to allow detailed comparison with measured
distributions~\cite{b:nlo_calc}. 
In this paper, we present a measurement of the 
dijet differential cross section that provides more precise information
about the initial state partons than has been probed by 
previous CDF measurements of inclusive jet transverse energy~\cite{b:inclusive}, total 
transverse energy~\cite{b:sumEt}, and dijet mass~\cite{b:dijetMass}. 
All previous measurements showed an excess of events at high jet energies when 
compared to the QCD prediction based on standard sets of parton 
distribution functions (PDFs). One explanation for this 
excess is a larger than expected number of high momentum partons,
particularly gluons, in the 
proton~\cite{b:cteq4hj,b:cteq4hjHighEt}. 
While those measurements
provide cross sections averaged over a wide range in their variable,
in this analysis we reduce the region over which averages are taken
by measuring the cross section for four separate ranges. This provides
more detailed information about the cross section shape.
Previous measurements  of the dijet differential 
cross section have been performed by the  CDF~\cite{b:dijetXsec1} and 
D$\O$~\cite{b:dijetXsecD0} collaborations with smaller data samples.
The present measurement places new constraints on the
parton distributions of the proton.

Jet production rates are usually expressed in terms of the transverse energy, 
$E_T$, and pseudorapidity, $\eta$, of the jets, where $\eta$ is related
to the polar angle $\theta$ relative to the proton beam line by 
$\eta \equiv$ -ln[tan($\theta/2$)]. 
At leading order in QCD, the proton, $p$,  and anti-proton, $\overline p$,  
momentum fractions, 
$x_1$ and $x_2$,
carried by the two colliding partons can be expressed as
\begin{equation}
\label{e:2jet}
x_{1}   =  \frac{E_T}{\sqrt{s}} (e^{\eta_1} + e^{\eta_2}),
\hspace{2cm}
x_{2}   =  \frac{E_T}{\sqrt{s}} (e^{-\eta_1} + e^{-\eta_2}).
\end{equation}
Here $\eta_1$ and $\eta_2$ are the pseudorapidities of the two
jets, $\sqrt{s}$ is the center of mass energy of the colliding
hadrons and $E_T$ is the transverse energy of the leading jet.
For a fixed $E_T$ and $\eta_1$, one can probe higher $x$ values
by selecting events in which the second jet has a larger $\eta_2$ value.
For a given $x$ we have four measurements at what are effectively 
different values of $Q^2$ the square of the four-momentum transfered in the 
interaction, calculated by
\begin{equation}
\label{e:q2}
Q^2 = 2 E_T^2 \cosh^2 \eta^* (1 - \tanh \eta^*),
\hspace{2cm}
\eta^* = \frac{1}{2} (\eta_1 - \eta_2).
\end{equation}
The four distributions in this analysis allow us 
to measure the cross section on a surface in the $x$-$Q^2$ phase space whose 
shape is sensitive to the predictions of different PDFs. 

The constraint on the parton distributions at high $x$ comes mainly from 
prompt photon production in $pp$ or $pA$ collisions from WA70~\cite{b:WA70} 
and the E706~\cite{b:E706} experiments and inclusive jet data from 
the Tevatron~\cite{b:inclusive}. The data do not constrain the 
parton distributions
very well at high $x$.
The higher statistics of this measurement together with the multiple
cross section measurements at different $Q^2$ for approximately the same 
$x$ provide a precise set of data which can be used to determine improved
sets of PDFs. 
The current measurement, based on data of an integrated 
luminosity of 86~${\rm pb}^{-1}$ from  
1.8~TeV~$\pbarp$ collisions taken during the 1994-1995 Fermilab Tevatron 
collider run, covers the range 
$0.05 \stackrel{<}{\sim} x_1 \stackrel{<}{\sim} 0.8$. 

The CDF detector is described in detail in~\cite{b:CDF_detector}.
In this analysis we 
utilize the central, plug, and forward calorimeters. The central
calorimeter covers the pseudorapidity range $|\eta| < 1.1$.
It is segmented into projective 
towers of size $\Delta\eta \times \Delta\phi =  0.1 \times 0.26$, 
where $\phi$ is
the azimuthal angle in radians. The plug
($1.1 < |\eta| < 2.4$) and forward ($2.4 < |\eta| < 4.2$) calorimeters are
segmented by
approximately 5$^\circ$ in $\phi$ and 0.1 in $\eta$. The event vertex is 
resolved to within 1~mm along the $z$ axis, using time projection 
chambers surrounding the beam pipe. 

A cone algorithm with cone radius
R $\equiv\sqrt{(\Delta\phi)^2 + (\Delta\eta)^2} = 0.7$ is used to 
identify jets~\cite{b:cdf_cone}. 
Transverse energy is defined as
$E_T = E \sin\theta$, where $E$ is the scalar sum of energy deposited 
in the calorimeter towers within the cone and $\theta$ is the angle 
formed by the event vertex, the beam direction, and the cone center.
Our data sample consists of events
collected by on-line identification of at least one jet
with transverse energy above trigger thresholds of 20, 50, 70, and 100~GeV
at integrated luminosities
of 0.091, 2.2, 11, and 86~${\rm pb}^{-1}$, respectively.
The bin widths in $E_T$ were chosen to be larger than the measurement 
resolution on $E_T$ and to ensure sufficient statistics in the bins.

In this analysis we use events with at least two jets 
of $E_T > 10$~GeV of uncorrected energy. 
We consider events in which the $E_T$-weighted centroid of
at least one of the two highest $E_T$ jets is in the range
$0.1 < | \eta | <0.7$.  This ``leading'' jet is required 
to deposit more than 40 GeV $E_T$, prior to corrections, 
in the central calorimeter.
In addition, the centroid of the second leading jet is required
to be in the region $0.1 < |\eta| < 3.0$, and
the primary event vertex must be located within $\pm 60$~cm of the 
nominal interaction point. 
Poorly measured events and background from cosmic rays, beam halo, and detector 
noise are removed by requiring 
that total energy recorded by the detector be less than 2000 GeV and 
$E\!\!\!\!/\,_T / \sqrt{\Sigma E_T} < 6 \sqrt{\rm GeV}$, where  
$E\!\!\!\!/\,_T$ is the missing transverse energy and $\Sigma E_T$
is the scalar sum of the total transverse energy. 

In this analysis, we evaluate
the $E_T$ spectrum of the leading jet for the following 
four $\eta$ bins of the second leading jet in the event:\\
\centerline{$0.1~<~|\eta_2|~<~0.7$, \hspace{2cm}  $0.7~<~|\eta_2|~<~1.4$,}\\
\centerline{$1.4~<~|\eta_2|~<~2.1$, \hspace{2cm}  $2.1~<~|\eta_2|~<~3.0$.}
The $\eta_2$ ranges  were chosen to place regions of reduced 
response (due to gaps between detectors) within single bins while 
at the same time maintaining a sufficient number of events in the bins. 
Both jets are included in the distribution for the 
$0.1~<~|\eta_2|~<~0.7$ bin
if each satisfies 
the requirement $0.1 < |\eta| < 0.7$ and $E_T > 40$ GeV.

Since the calorimetric response varies as a function of $\eta$, 
we determine the trigger response
separately for each $\eta_2$ bin.
The trigger efficiency was measured using overlapping $E_T$ regions
for the different trigger thresholds. For the  20 GeV trigger threshold, 
for which no lower $E_T$ trigger was available, the second jet in the 
event was used to determine the trigger efficiency. 
For the four trigger thresholds, the trigger 
efficiency was found to be greater than 90\% for jets of $E_T$ greater 
than 40, 82, 105, and 130 GeV.

\begin{figure}[h]
\centerline{
\psfig{figure=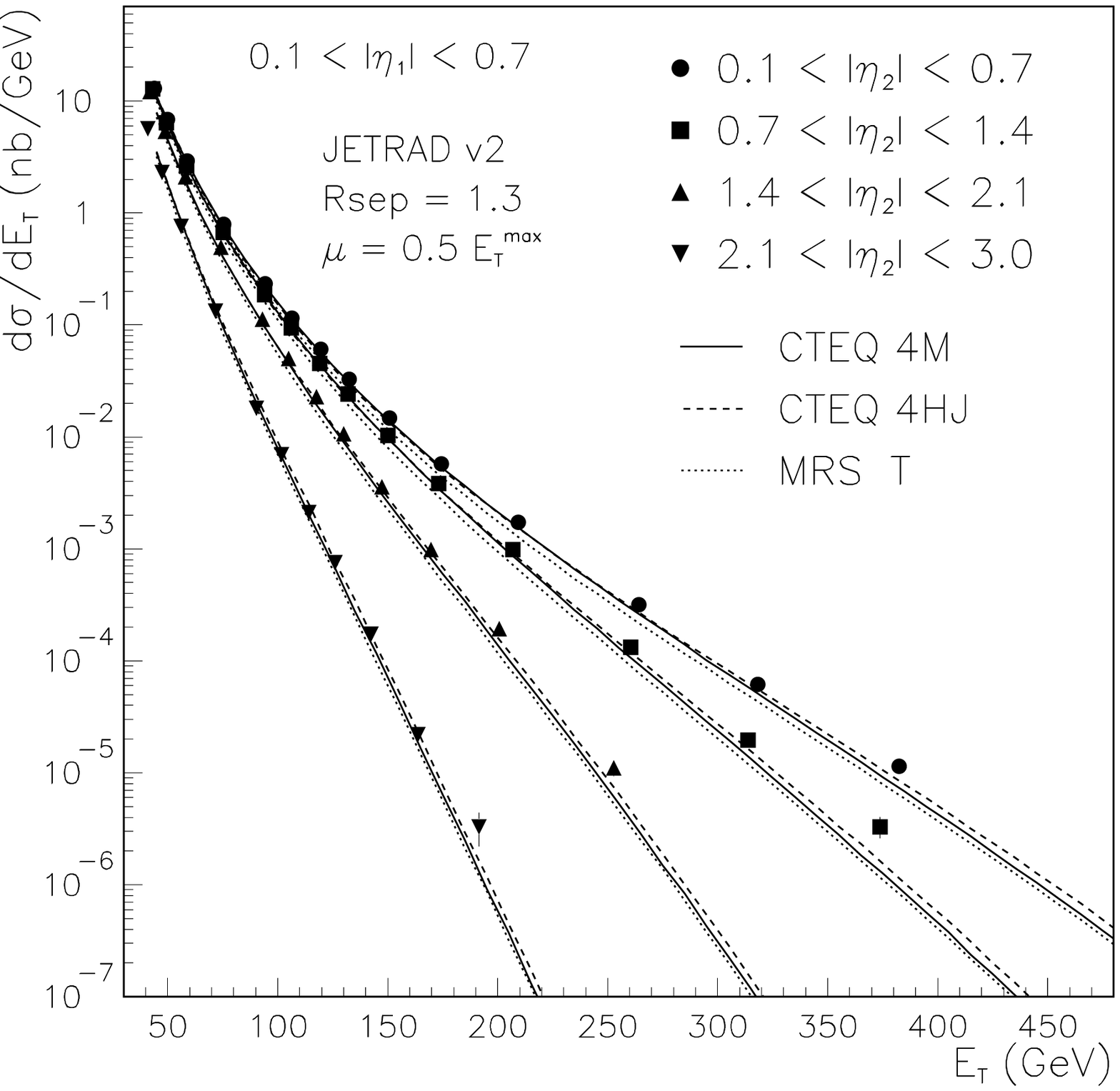,width=16cm}
}
\caption{\label{f:1plot}
The differential cross section for dijet events as a function of
transverse energy, $E_T$, and pseudorapidity, $\eta$, of one jet, for 4 ranges in
the pseudorapidity of the other jet.
The results are compared with QCD predictions using
different parton distribution functions. 
}
\end{figure}

The measured jet $E_T$ must be corrected for 
calorimeter non-linearity and loss of
energy in the gaps between calorimeters. In addition, the measured 
jet $E_T$ spectrum must be corrected for the smearing effect caused
by the resolution in the measured jet $E_T$. 
We simultaneously correct all these effects 
with the procedure used in our previous measurement of the 
inclusive jet $E_T$ spectrum~\cite{b:inclusive}.
For the central $\eta$ bin ($0.1 < |\eta_1| < 0.7$) at 40 GeV, 
the correction to the 
measured $E_T$ is approximately 4\%, while the correction to the 
measured cross section is about 19\%. The correction to the cross section 
increases to 70\% for the bin $2.1 < |\eta_1| < 3.0$.
The corrected cross section values are given in Tables~\ref{t:sigmaX1} 
and~\ref{t:sigmaX2} and plotted in Figure~\ref{f:1plot}.

\begin{table}
\centering
\begin{tabular}{|c|c|c|c||c|c|c|c|}       \hline
\multicolumn{4}{|c||}{$0.1 < |\eta_2| < 0.7$} & \multicolumn{4}{|c|}{$0.7 < |\eta_2| < 1.4$}  \\ \hline
$<E_T>$ & $d\sigma/dE_T$ & stat & sys  & $<E_T>$ & $d\sigma/dE_T$  & stat & sys      \\ 
 (GeV)  & (nb/GeV)       &  \%  &  \%  &  (GeV)  & (nb/GeV)        &  \%  &  \%      \\ \hline \hline
  44.0 &  1.23 $\times 10^{  1}$ &  1.3 & 19.5 &    43.1 &  1.29 $\times 10^{  1}$ &  1.3 & 21.5 \\ 
  50.0 &  6.48 $\times 10^{  0}$ &  1.7 & 18.4 &    49.5 &  6.41 $\times 10^{  0}$ &  1.7 & 19.6 \\ 
  58.9 &  2.78 $\times 10^{  0}$ &  1.6 & 17.2 &    58.5 &  2.65 $\times 10^{  0}$ &  1.6 & 17.8 \\ 
  75.5 &  7.54 $\times 10^{ -1}$ &  2.4 & 15.9 &    75.2 &  6.73 $\times 10^{ -1}$ &  2.5 & 16.2 \\ 
  94.3 &  2.22 $\times 10^{ -1}$ &  1.2 & 15.3 &    94.0 &  1.87 $\times 10^{ -1}$ &  1.3 & 15.7 \\ 
 106.6 &  1.10 $\times 10^{ -1}$ &  1.5 & 14.6 &   106.2 &  9.36 $\times 10^{ -2}$ &  1.6 & 15.1 \\ 
 119.5 &  5.83 $\times 10^{ -2}$ &  1.1 & 14.6 &   119.1 &  4.54 $\times 10^{ -2}$ &  1.2 & 15.2 \\ 
 132.5 &  3.13 $\times 10^{ -2}$ &  1.2 & 14.5 &   132.0 &  2.41 $\times 10^{ -2}$ &  1.3 & 15.4 \\ 
 150.8 &  1.42 $\times 10^{ -2}$ &  0.6 & 14.8 &   150.0 &  1.03 $\times 10^{ -2}$ &  0.6 & 16.0 \\ 
 174.4 &  5.53 $\times 10^{ -3}$ &  0.8 & 15.2 &   173.2 &  3.85 $\times 10^{ -3}$ &  0.9 & 16.9 \\ 
 209.4 &  1.67 $\times 10^{ -3}$ &  1.1 & 16.2 &   206.9 &  9.92 $\times 10^{ -4}$ &  1.4 & 18.9 \\ 
 264.1 &  3.10 $\times 10^{ -4}$ &  2.5 & 18.3 &   260.5 &  1.33 $\times 10^{ -4}$ &  3.9 & 22.9 \\ 
 318.2 &  6.06 $\times 10^{ -5}$ &  5.9 & 20.7 &   313.7 &  1.98 $\times 10^{ -5}$ & 10.4 & 27.7 \\ 
 382.4 &  1.14 $\times 10^{ -5}$ & 10.9 & 24.5 &   373.9 &  3.37 $\times 10^{ -6}$ & 21.3 & 34.6 \\ \hline
\end{tabular}
\caption{\label{t:sigmaX1}
The measured dijet differential cross sections for $0.1 < |\eta_2| < 0.7$ and
$0.7 < |\eta_2| < 1.4$. The differential cross section is given for the 
average $E_T$ of the bin. The statistical and systematic errors are shown 
as a percentage of the central value. 
}
\end{table}

\begin{table}
\centering
\begin{tabular}{|c|c|c|c||c|c|c|c|}       \hline
\multicolumn{4}{|c||}{$1.4 < |\eta_2| < 2.1$} & \multicolumn{4}{|c|}{$2.1 < |\eta_2| < 3.0$}  \\ \hline
$<E_T>$ & $d\sigma/dE_T$ & stat & sys  & $<E_T>$ & $d\sigma/dE_T$  & stat & sys      \\ 
 (GeV)  & (nb/GeV)       &  \%  &  \%  &  (GeV)  & (nb/GeV)        &  \%  &  \%      \\ \hline \hline
  42.1 &  1.14 $\times 10^{  1}$ &  1.4 & 22.6 &    40.9 &  5.81 $\times 10^{  0}$ &  2.0 & 27.3 \\ 
  48.9 &  5.04 $\times 10^{  0}$ &  1.9 & 20.8 &    47.5 &  2.39 $\times 10^{  0}$ &  2.7 & 25.1 \\ 
  58.0 &  1.99 $\times 10^{  0}$ &  1.9 & 19.4 &    56.2 &  7.85 $\times 10^{ -1}$ &  3.0 & 23.7 \\ 
  74.3 &  4.72 $\times 10^{ -1}$ &  3.0 & 18.4 &    71.7 &  1.40 $\times 10^{ -1}$ &  5.8 & 23.5 \\ 
  93.0 &  1.08 $\times 10^{ -1}$ &  1.7 & 18.5 &    90.4 &  1.93 $\times 10^{ -2}$ &  4.3 & 25.0 \\ 
 104.9 &  4.83 $\times 10^{ -2}$ &  2.3 & 18.4 &   101.8 &  7.47 $\times 10^{ -3}$ &  6.1 & 26.1 \\ 
 117.5 &  2.21 $\times 10^{ -2}$ &  1.8 & 19.0 &   114.2 &  2.28 $\times 10^{ -3}$ &  5.9 & 27.9 \\ 
 130.0 &  1.03 $\times 10^{ -2}$ &  2.1 & 19.7 &   126.0 &  8.13 $\times 10^{ -4}$ &  8.3 & 29.8 \\ 
 147.4 &  3.48 $\times 10^{ -3}$ &  1.2 & 21.0 &   142.5 &  1.89 $\times 10^{ -4}$ &  5.5 & 32.7 \\ 
 169.8 &  9.59 $\times 10^{ -4}$ &  2.0 & 22.9 &   163.7 &  2.39 $\times 10^{ -5}$ & 14.7 & 36.4 \\ 
 200.8 &  1.88 $\times 10^{ -4}$ &  3.6 & 26.3 &   191.4 &  3.64 $\times 10^{ -6}$ & 33.3 & 40.4 \\ 
 252.7 &  1.06 $\times 10^{ -5}$ & 15.8 & 33.3 &         &                         &       &     \\ \hline 
\end{tabular}
\caption{\label{t:sigmaX2}
The measured dijet differential cross sections for $1.4 < |\eta_2| < 2.1$ and
$2.1 < |\eta_2| < 3.0$.
The differential cross section is given for the average $E_T$ of the bin. 
The statistical and systematic errors are shown as a percentage of the central value.
}
\end{table}

The systematic error on the measurement of the jet cross section is dominated by
the uncertainty in the measurement of the jet $E_T$ magnified by the steep slope
of the $E_T$
spectrum. Although the same sources of uncertainty contribute to 
the cross section of each $E_T$ bin, the uncertainty depends on the local 
slope of the $E_T$ spectrum. The systematic uncertainties were evaluated as in 
References~\cite{b:inclusive} and~\cite{b:incSysErrCorrection}.
The uncertainties include: 
charged hadron response at high $p_T$ (h pt);
calorimeter response to low-$p_T$ hadrons (l pt);
$\pm$ 1 \% on the jet energy of the absolute calibration of the 
calorimeter (esc);
jet fragmentation functions used in the simulation (frag);
$\pm$ 30\% on the underlying event energy in the jet cone (uevt);
detector response to electrons and photons (e/ph);
and modeling of the detector jet energy resolution (cres).
The resolution on the measured $\eta$ causes events to migrate between
adjacent bins. In the highest $\eta$ bin, the gap between
the plug and forward calorimeters results in decreased $\eta$ resolution
and has the effect that more events migrate out of the bin than into it. 
To compensate for this effect, we have applied an $E_T$-dependent 
correction which is less than 8\% in all
bins. The effect was studied by breaking it into two components,
the resolution on the measured $\eta$ ($\eta$ res)
and
a systematic shift in the reconstructed $\eta$ ($\eta$ sh).
It is included in the systematic error by looking at 
the result on the cross section when doubling and halving the correction.  
Bins for which events were collected using triggers with uncorrected 
energy greater than 20 GeV (J20), 50 GeV (J50) 
and 70 GeV (J70) were assigned 4, 2 and 2 percent errors 
respectively, associated with prescaling.
An overall luminosity uncertainty (norm) of 4 percent is added in 
quadrature with these. 
The sources of systematic errors are listed in 
Tables~\ref{t:sysErrors1} through~\ref{t:sysErrors4} as percentages of the 
central $E_T$ value for each $E_T$ and $\eta$  bin. 
In general the percent error increases as $\eta_2$ increases.

\begin{table}
\centering
\centerline{
\small
\begin{tabular}{|c||c|c|c|c|c|c|c|c|c|c|c|c|c|c|}       \hline
 $<\!E_T\!>$  &  e/ph &  uevt &  frag &  esc &  cres &  l pt &  h pt &  $\eta$ sh &
 $\eta$ res & norm & J20 & J50 & J70 & tot \\ \hline \hline
  43.9 &  2.5 & 12.7 &  7.9 &  4.1 &  4.9 &  7.3 &  2.6 &  2.5 &  0.4 &  4.0 &  4.0 &  2.0 &  2.0 & 19.5\\
  50.0 &  2.6 & 11.1 &  8.0 &  4.0 &  4.5 &  7.3 &  2.9 &  2.5 &  0.4 &  4.0 &  4.0 &  2.0 &  2.0 & 18.4\\
  58.9 &  2.6 &  9.1 &  8.1 &  4.0 &  4.0 &  7.1 &  3.4 &  2.5 &  0.4 &  4.0 &  4.0 &  2.0 &  2.0 & 17.2\\
  75.5 &  2.8 &  6.6 &  8.1 &  4.1 &  3.2 &  6.7 &  4.3 &  2.5 &  0.4 &  4.0 &  4.0 &  2.0 &  2.0 & 15.9\\
  94.3 &  2.9 &  5.1 &  7.9 &  4.3 &  2.6 &  6.2 &  5.3 &  2.4 &  0.4 &  4.0 &  4.0 &  2.0 &  2.0 & 15.3\\
 106.6 &  2.9 &  4.4 &  7.7 &  4.5 &  2.3 &  5.9 &  5.9 &  2.4 &  0.4 &  4.0 &  0.0 &  2.0 &  2.0 & 14.6\\
 119.5 &  3.0 &  4.0 &  7.6 &  4.6 &  2.1 &  5.5 &  6.6 &  2.4 &  0.4 &  4.0 &  0.0 &  2.0 &  2.0 & 14.6\\
 132.5 &  3.1 &  3.6 &  7.4 &  4.8 &  2.0 &  5.2 &  7.3 &  2.3 &  0.4 &  4.0 &  0.0 &  0.0 &  2.0 & 14.5\\
 150.8 &  3.2 &  3.3 &  7.3 &  5.1 &  2.0 &  4.8 &  8.1 &  2.3 &  0.4 &  4.0 &  0.0 &  0.0 &  2.0 & 14.8\\
 174.4 &  3.3 &  3.1 &  7.2 &  5.4 &  2.1 &  4.4 &  9.1 &  2.2 &  0.4 &  4.0 &  0.0 &  0.0 &  0.0 & 15.2\\
 209.4 &  3.6 &  2.9 &  7.4 &  5.9 &  2.4 &  4.0 & 10.5 &  2.0 &  0.4 &  4.0 &  0.0 &  0.0 &  0.0 & 16.2\\
 264.1 &  4.2 &  2.7 &  8.3 &  6.7 &  2.9 &  4.0 & 12.3 &  1.7 &  0.3 &  4.0 &  0.0 &  0.0 &  0.0 & 18.3\\
 318.2 &  4.9 &  2.5 & 10.0 &  7.4 &  3.5 &  4.8 & 13.8 &  1.2 &  0.3 &  4.0 &  0.0 &  0.0 &  0.0 & 20.7\\
 382.4 &  5.8 &  2.3 & 13.4 &  8.2 &  5.0 &  6.8 & 15.0 &  0.6 &  0.9 &  4.0 &  0.0 &  0.0 &  0.0 & 24.5\\ \hline
\end{tabular}
}
\caption{\label{t:sysErrors1}
The systematic errors for the $0.1 < \eta_2 < 0.7$ bin given as 
a percentage of the central value. The $E_T$ values
are specified at the bin average. A reference to the sources 
of systematic  errors is given in the text.
}
\end{table}

\begin{table}
\centering
\centerline{
\small
\begin{tabular}{|c||c|c|c|c|c|c|c|c|c|c|c|c|c|c|}       \hline
 $<\!E_T\!>$  &  e/ph &  uevt &  frag &  esc &  cres &  l pt &  h pt &  $\eta$ sh &
 $\eta$ res & norm & J20 & J50 & J70 & tot \\ \hline \hline
  43.1 &  2.6 & 15.0 &  8.3 &  4.2 &  6.1 &  7.7 &  2.7 &  0.6 &  0.3 &  4.0 &  4.0 &  2.0 &  2.0 & 21.5\\
  49.5 &  2.6 & 12.3 &  8.5 &  4.1 &  5.4 &  7.6 &  3.1 &  0.5 &  0.3 &  4.0 &  4.0 &  2.0 &  2.0 & 19.6\\
  58.5 &  2.7 &  9.5 &  8.6 &  4.1 &  4.6 &  7.4 &  3.6 &  0.5 &  0.3 &  4.0 &  4.0 &  2.0 &  2.0 & 17.8\\
  75.2 &  2.8 &  6.6 &  8.5 &  4.1 &  3.4 &  7.0 &  4.6 &  0.5 &  0.3 &  4.0 &  4.0 &  2.0 &  2.0 & 16.2\\
  94.0 &  3.0 &  5.0 &  8.4 &  4.3 &  2.6 &  6.5 &  5.7 &  0.4 &  0.2 &  4.0 &  4.0 &  2.0 &  2.0 & 15.7\\
 106.2 &  3.1 &  4.5 &  8.3 &  4.5 &  2.3 &  6.2 &  6.4 &  0.5 &  0.2 &  4.0 &  0.0 &  2.0 &  2.0 & 15.1\\
 119.1 &  3.2 &  4.1 &  8.2 &  4.8 &  2.2 &  5.9 &  7.2 &  0.5 &  0.2 &  4.0 &  0.0 &  2.0 &  2.0 & 15.2\\
 131.9 &  3.3 &  3.9 &  8.1 &  5.1 &  2.1 &  5.6 &  7.9 &  0.5 &  0.2 &  4.0 &  0.0 &  0.0 &  2.0 & 15.4\\
 150.0 &  3.5 &  3.8 &  8.1 &  5.5 &  2.3 &  5.2 &  9.0 &  0.6 &  0.3 &  4.0 &  0.0 &  0.0 &  2.0 & 16.0\\
 173.2 &  3.8 &  3.7 &  8.2 &  6.1 &  2.6 &  4.9 & 10.3 &  0.7 &  0.3 &  4.0 &  0.0 &  0.0 &  0.0 & 16.9\\
 206.9 &  4.2 &  3.6 &  8.7 &  7.1 &  3.3 &  4.7 & 12.2 &  1.0 &  0.5 &  4.0 &  0.0 &  0.0 &  0.0 & 18.9\\
 260.5 &  5.1 &  3.5 & 10.4 &  8.9 &  4.5 &  5.0 & 15.2 &  1.7 &  1.1 &  4.0 &  0.0 &  0.0 &  0.0 & 22.9\\
 313.7 &  6.0 &  3.4 & 13.2 & 10.9 &  5.4 &  6.0 & 18.2 &  2.6 &  2.3 &  4.0 &  0.0 &  0.0 &  0.0 & 27.7\\
 373.9 &  7.1 &  3.2 & 17.9 & 13.5 &  6.7 &  8.1 & 21.7 &  3.9 &  4.7 &  4.0 &  0.0 &  0.0 &  0.0 & 34.6\\ \hline
\end{tabular}
}
\caption{\label{t:sysErrors2}
The systematic errors for the $0.7 < \eta_2 < 1.4$ bin given as a 
percentage of the central value. The $E_T$ values
are specified at the bin average. A reference to the sources 
of systematic  errors is given in the text.
}
\end{table}

\begin{table}
\centering
\centerline{
\small
\begin{tabular}{|c||c|c|c|c|c|c|c|c|c|c|c|c|c|c|}       \hline
 $<\!E_T\!>$  &  e/ph &  uevt &  frag &  esc &  cres &  l pt &  h pt &  $\eta$ sh &
 $\eta$ res & norm & J20 & J50 & J70 & tot \\ \hline \hline
  42.1 &  2.8 & 15.7 &  8.5 &  4.3 &  6.6 &  7.8 &  2.8 &  3.6 &  0.5 &  4.0 &  4.0 &  2.0 &  2.0 & 22.6\\
  48.9 &  2.9 & 12.8 &  8.9 &  4.6 &  5.8 &  8.0 &  3.1 &  3.4 &  0.5 &  4.0 &  4.0 &  2.0 &  2.0 & 20.8\\
  58.0 &  3.0 & 10.2 &  9.3 &  4.8 &  5.0 &  8.0 &  3.7 &  3.2 &  0.4 &  4.0 &  4.0 &  2.0 &  2.0 & 19.4\\
  74.3 &  3.2 &  7.6 &  9.6 &  5.1 &  4.0 &  7.9 &  5.1 &  3.0 &  0.3 &  4.0 &  4.0 &  2.0 &  2.0 & 18.4\\
  93.0 &  3.5 &  6.3 &  9.8 &  5.4 &  3.6 &  7.5 &  6.9 &  2.6 &  0.2 &  4.0 &  4.0 &  2.0 &  2.0 & 18.5\\
 104.9 &  3.7 &  5.8 & 10.0 &  5.6 &  3.5 &  7.3 &  8.0 &  2.4 &  0.2 &  4.0 &  0.0 &  2.0 &  2.0 & 18.4\\
 117.5 &  4.0 &  5.6 & 10.1 &  5.9 &  3.6 &  7.0 &  9.3 &  2.2 &  0.1 &  4.0 &  0.0 &  2.0 &  2.0 & 19.0\\
 130.0 &  4.2 &  5.4 & 10.3 &  6.3 &  3.8 &  6.9 & 10.5 &  2.1 &  0.1 &  4.0 &  0.0 &  0.0 &  2.0 & 19.7\\
 147.4 &  4.6 &  5.2 & 10.6 &  6.9 &  4.3 &  6.7 & 12.1 &  1.8 &  0.5 &  4.0 &  0.0 &  0.0 &  2.0 & 21.0\\
 169.7 &  5.1 &  5.1 & 11.2 &  7.8 &  5.0 &  6.5 & 14.1 &  1.6 &  1.2 &  4.0 &  0.0 &  0.0 &  0.0 & 22.9\\
 200.8 &  5.8 &  4.9 & 12.4 &  9.5 &  6.1 &  6.7 & 16.8 &  1.2 &  2.7 &  4.0 &  0.0 &  0.0 &  0.0 & 26.2\\
 252.7 &  7.3 &  4.6 & 15.4 & 13.2 &  8.5 &  7.8 & 20.8 &  0.8 &  6.7 &  4.0 &  0.0 &  0.0 &  0.0 & 33.3\\ \hline
\end{tabular}
}
\caption{\label{t:sysErrors3}
The systematic errors for the $1.4 < \eta_2 < 2.1$ bin given as a 
percentage. The $E_T$ values
are specified at the bin average. A reference to the sources 
of systematic  errors is given in the text.
}
\end{table}

\begin{table}
\centering
\centerline{
\small
\begin{tabular}{|c||c|c|c|c|c|c|c|c|c|c|c|c|c|c|}       \hline
\small
 $<\!E_T\!>$  &  e/ph &  uevt &  frag &  esc &  cres &  l pt &  h pt &  $\eta$ sh &
 $\eta$ res & norm & J20 & J50 & J70 & tot \\ \hline \hline
  40.9 &  3.1 & 19.9 &  9.7 &  4.9 &  9.0 &  9.0 &  3.2 &  3.1 &  0.5 &  4.0 &  4.0 &  2.0 &  2.0 & 27.3\\
  47.5 &  3.3 & 16.3 & 10.3 &  5.3 &  8.3 &  9.2 &  3.9 &  2.9 &  0.4 &  4.0 &  4.0 &  2.0 &  2.0 & 25.1\\
  56.2 &  3.6 & 13.4 & 10.9 &  5.7 &  7.5 &  9.5 &  4.8 &  2.7 &  0.2 &  4.0 &  4.0 &  2.0 &  2.0 & 23.7\\
  71.7 &  4.1 & 10.8 & 12.1 &  6.4 &  6.5 &  9.8 &  6.8 &  2.5 &  0.2 &  4.0 &  4.0 &  2.0 &  2.0 & 23.5\\
  90.4 &  4.8 &  9.7 & 13.3 &  7.4 &  6.5 & 10.0 &  9.3 &  2.9 &  0.7 &  4.0 &  4.0 &  2.0 &  2.0 & 25.0\\
 101.8 &  5.2 &  9.4 & 13.9 &  8.0 &  7.0 & 10.1 & 11.0 &  3.3 &  1.0 &  4.0 &  0.0 &  2.0 &  2.0 & 26.1\\
 114.2 &  5.7 &  9.2 & 14.5 &  8.7 &  8.0 & 10.1 & 12.9 &  4.0 &  1.3 &  4.0 &  0.0 &  2.0 &  2.0 & 27.9\\
 126.0 &  6.1 &  9.1 & 15.1 &  9.4 &  9.3 & 10.1 & 14.7 &  4.9 &  1.7 &  4.0 &  0.0 &  0.0 &  2.0 & 29.8\\
 142.4 &  6.8 &  9.0 & 15.8 & 10.3 & 11.0 & 10.0 & 17.4 &  6.4 &  2.2 &  4.0 &  0.0 &  0.0 &  2.0 & 32.7\\
 163.6 &  7.7 &  8.9 & 16.5 & 11.5 & 11.9 &  9.7 & 20.9 &  8.9 &  2.9 &  4.0 &  0.0 &  0.0 &  0.0 & 36.4\\
 191.4 &  8.8 &  8.8 & 17.3 & 13.0 &  7.3 &  9.3 & 25.8 & 13.2 &  4.0 &  4.0 &  0.0 &  0.0 &  0.0 & 40.4\\ \hline
\end{tabular}
}
\caption{\label{t:sysErrors4}
The systematic errors for the $2.1 < \eta_2 < 3.0$ bin given as a 
percentage. The $E_T$ values
are specified at the bin average. A reference to the sources 
of systematic  errors is given in the text.
}
\end{table}

\begin{figure}[h]
\centerline{
\psfig{figure=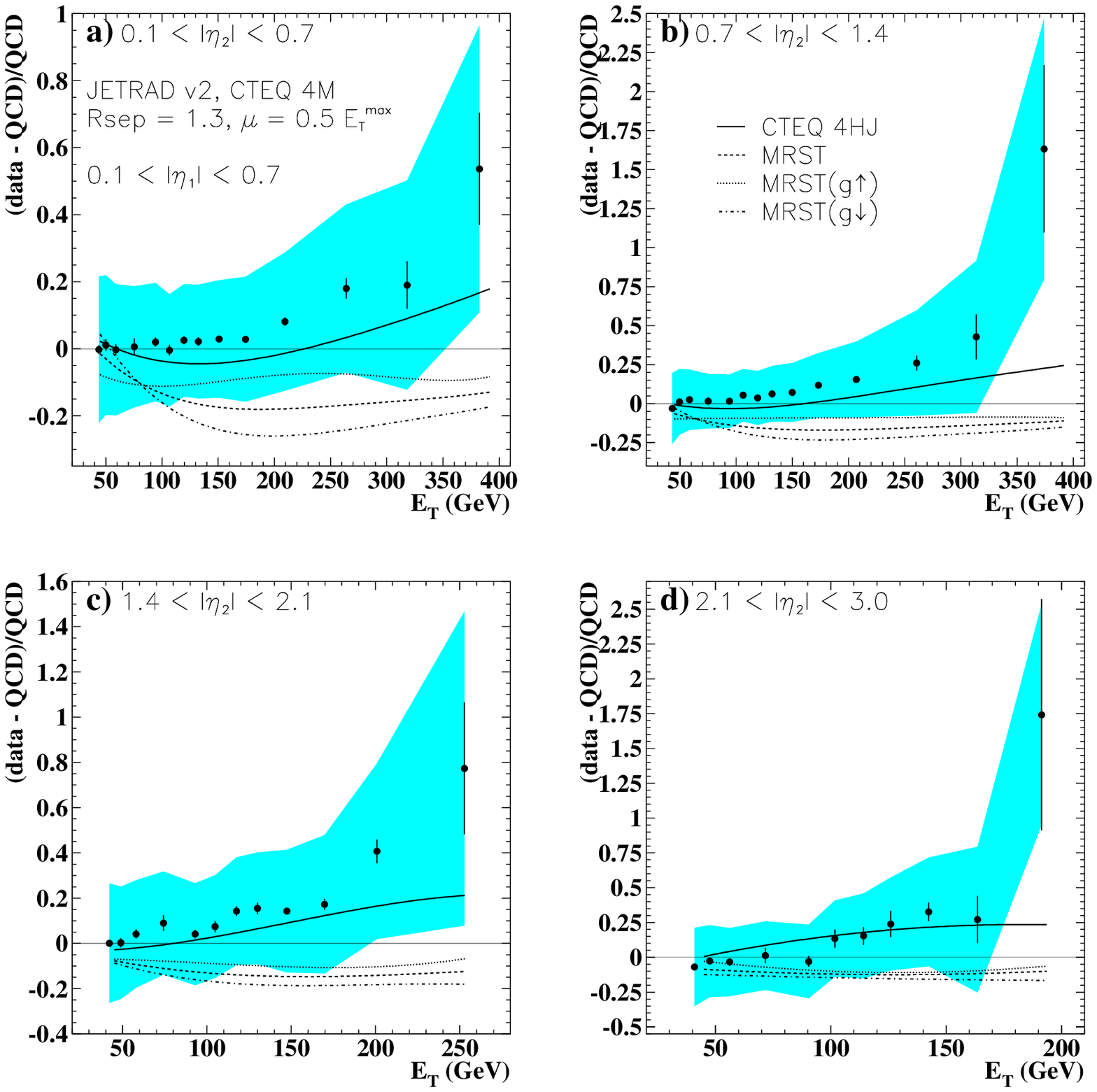,width=16cm}
}
\caption{\label{f:dt}
The differential cross section for dijet events as a function of
transverse energy, $E_T$, and pseudorapidity, $\eta$, of the leading jet, 
for four ranges in
the pseudorapidity of the second leading jet.
The results are compared with QCD predictions using
different parton distribution functions. 
The statistical error is represented by the error bars while the 
correlated systematic error is shown as the shaded band.
}
\end{figure}

In Figure~\ref{f:dt}, the difference between the fully corrected two-jet 
differential cross section and the predicted cross section is divided 
by the predicted cross section and plotted as a function of the leading 
jet $E_T$ for the four $\eta$ ranges of the second jet.
The theory predictions were calculated
using the NLO calculation of the JETRAD program~\cite{b:jetrad} with the
PDFs indicated.
The calculations use a renormalization 
scale $\mu = E_T^{max}/2$ with ${\rm R}_{sep} = 1.3$, 
where ${\rm R}_{sep}$ is a measure of the maximum separation
between the cones of two jets that are merged into one.
The error bars represent the statistical errors, while the
shaded bands represent one standard deviation of the systematic error,
which is correlated for all the different $E_T$ values.
The data are compared to the predicted cross section obtained 
using the PDF set CTEQ4M~\cite{b:cteq4hj}. 
The solid curve shows the expected results when using 
CTEQ4HJ~\cite{b:cteq4hj}, and the dashed curves show the results when 
using the PDF set MRST~\cite{b:mrst}.

The observed excess of events at high $E_T$ values in the 
inclusive jet cross section measurement may be explained 
within the framework of conventional 
QCD by exploiting the relatively weak restriction on the gluon density at
high $x$~\cite{b:cteq4hjHighEt}. The CTEQ4 PDFs use a more flexible 
parameterization of the gluon density at high $x$ 
than is present in other sets. The CTEQ4 set of PDFs include the
inclusive jet data from the Tevatron.
The CTEQ4HJ PDF gives a higher weight to the inclusive jet data
while still maintaining agreement with the other data sets used in the fit. 

The MRST set of PDFs is based in a wide range of deep inelastic
scattering data and has an improved treatment of heavy flavors
and prompt photon production than do previous MRST sets. 
The main constraint upon the gluon at high $x$ comes from prompt photon 
production from the WA70~\cite{b:WA70} and E706~\cite{b:E706} data. 
The set MRST(g$\uparrow$) was derived assuming that there is no
initial state partonic transverse momentum ($<\!k_T\!>\,=0$); this
does not lead to a good fit for the prompt photon data from the 
E706 experiment.  
The set labelled MRST(g$\downarrow$) was derived by allowing 
non-zero $<\!k_T\!>$  while maintaining reasonable agreement
with the WA70 data. The  MRST(g$\downarrow$) set has {$<\!k_T\!>\,=0.64\,{\rm GeV}$}.
These two sets represent the extreme values of $<\!k_T\!>$ that yield 
reasonable agreement with the data used in the fit. 
The set labelled MRST represents the preferred set from the
global analysis and has $<\!k_T\!> \, = 0.4 \, {\rm GeV}$.

The covariance matrix for the dijet cross section is 
\[
V_{ij} = \delta_{ij} \sigma^2_i(stat) + \Sigma_{k=1}^{13} \sigma_i(sys_k) \sigma_j(sys_k),
\]
where
$\delta_{ij} = 1(0)$
for $i = j(i \ne j)$ $\sigma_i(stat)$ is the statistical uncertainty 
in bin $i$ and $\sigma_i(sys_k)$ is the systematic uncertainty, $k$, on bin 
$i$. The sum is over the 13 sources of systematic errors listed above and 
over all the $E_T$ bins in each of the four $\eta$ bins.  We 
calculate the $\chi^2$ from $\chi^2 = \Sigma_{ij} \Delta_i (V^{-1})_{ij} \Delta_j$,
where $\Delta_i$ is the difference between the data and theoretical 
prediction for bin $i$. 
The average of the upper and lower errors is used when calculating the 
$\chi^2$. The $\chi^2/{\rm dof}$ values for different PDFs are presented 
in Table~\ref{t:totalchi}.
Although the cross sections predicted by the MRST PDFs are lower than the data by 20\%, they 
have similar $\chi^2$ values to those predicted with CTEQ4M. This is because the systematic 
errors allow a correlated shift in the data which makes only a small contribution 
to the total $\chi^2$. Predictions whose shape matches that of a correlated
systematic error will give reasonable $\chi^2$ values provided that the normalization 
between the data and prediction are within a few standard deviations. 
The probability of describing the data with the PDFs used in this analysis is
less than 1\% in all cases. 

\begin{table}[h!]
\centering
\begin{tabular}{|l|l||l|l|}
\hline
PDF                 & $\chi^2$/dof & PDF     & $\chi^2$/dof  \\ \hline
MRST                & 2.68         & CTEQ4HJ & 2.43          \\
MRST(g$\uparrow$)   & 3.63         & CTEQ4M  & 2.88          \\  
MRST(g$\downarrow$) & 4.49         &         &               \\ \hline
\end{tabular}
\caption{\label{t:totalchi}
The $\chi^2$/dof between the data and the prediction of Giele, et al., 
for different PDFs. The fit to the data has 51 degrees of freedom.
}
\end{table}

In summary, we have measured the differential cross section for dijet 
production in $\pbarp$ collisions with one jet restricted to the 
pseudorapidity region $0.1 < |\eta_1| < 0.7$ for four different 
pseudorapidity bins of a second jet restricted within $0.1 < |\eta_2| < 3.0$.
By allowing the pseudorapidity of the second jet to vary through
$0.1 < |\eta| < 3.0$, we are able to 
map out the cross section over the available kinematic phase space
and provide a differential cross section that more tightly constrains
the parton distributions of the proton than in measurements 
previously reported by us.
The measurement provides more precise information about 
the parton distributions of the proton in the high $x$ region, an area which 
is not well constrained, and will provide useful input to QCD global fits.
The resulting improved sets of PDFs will help to further 
enhance our knowledge of the structure functions of the proton.

We thank the Fermilab staff and the technical staffs of the
participating institutions for their vital contributions.  This work was
supported by the U.S. Department of Energy and National Science Foundation;
the Italian Istituto Nazionale di Fisica Nucleare; the Ministry of Education,
Science, Sports and Culture of Japan; the Natural Sciences and Engineering 
Research Council of Canada; the National Science Council of the Republic of 
China; the Swiss National Science Foundation; the A. P. Sloan Foundation; the
Bundesministerium fuer Bildung und Forschung, Germany; and the Korea Science 
and Engineering Foundation.


\end{document}